\documentclass{bmcart}
\usepackage{siunitx,xr-hyper,hyperref,lipsum}
\usepackage[T1]{fontenc}

\usepackage{amsmath, amsthm, amssymb}
\usepackage[linesnumbered,ruled,vlined]{algorithm2e}
\usepackage{graphicx} 
\usepackage[caption=false,font=footnotesize]{subfig}
\usepackage{mathtools}
\newcommand{\mbf}[1]{\mathbf{#1}}     \newcommand{\mbs}[1]{\boldsymbol{#1}}     
\newcommand{\mb}[1]{\mbox{#1}}
\newcommand{\WRX}{\omega_{\mbox{\scriptsize rx}}}
\newcommand{\ttoa}{\tau_{\mbox{\scriptsize toa}}}
\newcommand{\tmax}{\tau_{\mbox{\scriptsize toa}}^{\scriptsize \mbox{max}}}
\newcommand{\httoa}{\hat{\tau}_{\mbox{\scriptsize toa}}}
\newcommand{\WTR}{\omega_{\mbox{\scriptsize tr}}}
\newcommand{\HH}{\mbs{\mathcal{H}}}
\newcommand{\pld}{\tau_{\mbox{\scriptsize pld}}}
\newcommand{\pldmax}{\tau_{\mbox{\scriptsize pld}}^{\scriptsize \mbox{max}}}
\newcommand{\bt}{\mbs{\theta}}
\newcommand{\be}[1]{\mbf{e}_{#1}}
\newcommand{\bh}[1]{\mbf{h}_{#1}}
\newcommand{\bb}[1]{\mbf{#1}}
\newcommand{\bs}[1]{\mbs{#1}}
\newcommand{\R}[1]{\mathbb{R}^{#1}}

\newcommand {\BC}{\mbs{\mathcal{C}}}
\newcommand {\BD}{\mbs{\mathcal{D}}}
\DeclareMathOperator*{\argmin}{arg\,min}     \DeclareMathOperator*{\argmax}{arg\,max}    
\DeclareMathAlphabet\mathbfcal{OMS}{cmsy}{b}{n}

\newcommand{\norm}[1]{\left\lVert#1\right\rVert}

\begin{document}

\begin{frontmatter}
\begin{fmbox}
\dochead{Research}
\title{Compressive Sampling Based UWB TOA Estimator }
\address[id=aff1]{
  \orgname{Department of Signal Processing, KTH Royal Institute of Technology}, 
  \street{Fack},                     %
  \postcode{100 44}                                
  \city{Stockholm},                              
  \cny{Sweden}                                    
}
\author[
   addressref={aff1},                   
   corref={aff1},                       
   email={vpy@kth.se}   
]{\inits{}\fnm{Vijaya} \snm{Yajnanarayana}}
\author[
   addressref={aff1},
   email={ph@kth.se}
]{\inits{}\fnm{Peter} \snm{H\"{a}ndel}}
\begin{artnotes}
\end{artnotes}

\end{fmbox}

\begin{abstractbox}
\begin{abstract}
This paper proposes two compressive sampling based time of arrival (TOA) estimation algorithms using a  sub-Nyquist  rate receiver. We also describe a novel compressive sampling dictionary design for the compact representation  of the received UWB signal. One of the proposed algorithm exploits the a-priori information with regard to the channel and range of the target. The performance of the  algorithms are compared against the maximum likelihood (ML) based receiver using IEEE 802.15.4a CM1 line of sight (LOS) UWB channel model. The proposed algorithm yields performance similar to  the ML TOA estimation at high SNRs. However, the computational complexity and the sampling rate requirements are lesser compared to the ML estimator. Simulation results show that the proposed algorithms can match ML estimator performance with only $1/4$-th the sampling rate at $25~\mb{dB}$ SNR. We analyze the performance of the algorithm with respect to practical constraints like size of the holographic dictionary and sampling rates.  We also propose a new algorithm which can exploit the a-priori information  regarding the UWB channel and the geographical constraints on the target that may be available at the receiver. This algorithm can substantially boost the performance compared to the algorithm without a-priori information at low SNRs.
\end{abstract}


\end{abstractbox}
%

\end{frontmatter}

\section{Introduction}
\label{sec:intro}

Compressive sampling (CS) technology has far reaching implications and concern  a number of varied applications such as data compression, channel coding, medical imaging, etc. New applications for this technology are emerging constantly. In this paper, we address a classical estimation problem concerning ultra-wideband (UWB) time of arrival (TOA) using compressive sampling technique. The compressive sampling theory suggests that from a fewer number of acquisition samples, which is less than that advocated by the Nyquist theory,  an approximate reconstruction of the original signal is possible. This involves choosing an appropriate measurement matrix for efficiently representing the received signal  in lower dimension \cite{Candes-sp-article,Hayashi-Tut}. The choice of reducing the sampling rate of the overall system is important as this enables digital designs for wideband systems like UWB.

An UWB system is based on spreading a low power signal into large bandwidth.  Impulse radio based UWB (IR-UWB) schemes are the most popular UWB technique. Narrow impulse like pulses used in IR-UWB schemes can yield very high time resolution and thus, can be used for accurate measurement of TOA \cite{Win1998,Witrisal}.

Availability of unlicensed frequency spectrum at frequency band  less than $10~\mb{GHz}$ created a lot of excitement for employing UWB for various applications including TOA. However, there are several challenges in adopting UWB at these frequency bands. The two main challenges include the stringent emission regulations by the regulatory bodies like federal communications commission (FCC) \cite{fcc2002, LV-CS}. This had a direct impact on the range and rate of the system.  Secondly, wide bandwidth requirement of UWB created bottleneck in the ADC design.

Availability of the mmWave spectrum without stringent emission requirements has benefited the evolution of UWB technology in the $30 -50~\mb{GHz}$ spectrum. This has improved the range and rate problems discussed earlier. However, large bandwidth  coupled with high time and amplitude resolution requirements for ADCs still persist and pose significant challenge in the design of the digital UWB system.

A digital UWB transceiver can offer flexibility and scalability. Moore's law gives us processing power for ``free'', and we can make savings by using a cheaper digital front-end. However, UWB signal occupies extremely wide bandwidth, thus requiring high sampling rate. For example,  TOA estimation using the proposed rake receiver structure  in \cite{Lottici}, would require sampling rates in excess of $25~\mb{GHz}$. The ADC design for wideband systems face  several challenges in order to support wide bandwidth which include among others amplitude resolution, sampling rate, analog bandwidth, cost, etc. \cite{RF-Charles,murmann-adc-survey}. High speed ADCs for UWB TOA systems can be designed using an interleaved flash ADC or bank of poly-phase ADCs \cite{flash-adc,polyphase-adc}. However, they are sensitive to timing jitter, their amplitude resolution is generally poor and are  expensive in terms of the cost. 

In this paper, we propose a sub-sampled UWB receiver based on compressive sampling. With IEEE 802.15.4a CM1 as the channel model \cite{uwb-channel-model,Molisch} and the Nyquist sampled maximum likelihood based  UWB TOA receiver as a starting point, we show that the requirements on RF front-end sampling rate can be significantly loosened by employing recent theories on compressive sampling, without any significant loss in performance at high SNR. To accomplished this, we will represent  the received UWB signal in a compact form using the columns of  a carefully chosen dictionary leading to a sparse signal representation.  We propose a TOA estimation method, which can estimate the TOA from this sparse representation. We also analyze the performance trade-off of the proposed algorithm in terms of the dictionary size, sampling rate and sparsity level. To the authors knowledge, compressive sampling has not before been applied to the considered problem, which is a main motivation for the paper. In addition, we also show that a-priori properties of the IEEE 802.15.4a channel model opens up for improvement of the receiver performance by taking statistical channel properties into account. This is a contribution that is believed to have an interest for the reader in its own right.

Reminder of this paper is organized as follows. In Section \ref{sec:survey}, we provide the background to TOA estimation and  an overview of state of the art TOA estimation methods. In Section \ref{sec:model}, we will discuss the received multipath signal. In Section \ref{sec:ML}, we will discuss the maximum likelihood (ML) based TOA estimation.  Section \ref{sec:channel}, introduces the UWB channels and discusses the challenges in estimating the TOA in multipath UWB channels. We use IEEE 802.15.4a CM1 line of sight (LOS) channel model as a reference. In Section \ref{sec:cs}, we briefly discuss the compressive sampling theory and present a method for sparse representation of the received UWB signal. This enables us to faithfully represent the received UWB signal at lower sampling  rate. Section \ref{sec:algo}, discusses two TOA estimation algorithms. Section~\ref{sec:res}, provides  the simulation results for TOA estimation  under different practical scenarios. Finally, Section~\ref{sec:con}, discusses the conclusions from the results presented.

\section{Background}
\label{sec:survey}
TOA estimation involves estimating the propagation time between transmitter and receiver. Accurate estimation of TOA is essential in several applications including positioning and communication.  In many communication systems, the message information is embedded in the location of the pulse. For example, TOA estimation  methods can be used  to demodulate pulse position modulated (PPM), IR-UWB symbols \cite{VJ-3,nikos-ppm}.  Another application of TOA is in the estimation of the position of mobile nodes in a wireless sensor network (WSN). Here, TOA values are used to calculate the range from the anchor nodes using which the location information can be derived \cite{JO-interagent,Rantakokko-2}.

There are several works concerning TOA estimation. Broadly, they can be classified as frequency-domain and time-domain methods. Frequency-domain methods,  typically involve estimating the frequency domain channel response by sweeping the channel using a multi-carrier modulation schemes such as OFDM. Then applying a super-resolution algorithm such as root multiple signal classification (MUSIC) \cite{MUSIC} or total least square-estimation of signal parameter via rotational invariance techniques (TLS-ESPRIT) \cite{TLS_ESPRIT}  on the channel frequency response \cite{Li-SuperResolution}. Due to the large number of multi-paths in the UWB propagation environment, the implementation complexity of this method is extremely high \cite{Falsi-toa}.

In the time-domain TOA estimation methods, the TOA is estimated directly from the received time-domain signal by identifying the first multipath. UWB based positioning systems uses wide bandwidth  and  provides high time resolution, therefore TOA estimation using time-based ranging is the method of choice in these systems. The simplest of these methods include choosing the location of the peak in the received signal as the TOA estimate \cite{gezici-book,Vidal-toa}. The main source of error in these TOA estimation methods is due to the strongest mutipath components arriving later than the first path. In UWB channels this can happen because each multipath component show delay dispersion by itself. That means a short pulse that, for example, undergoes only a single diffraction may arrive at the receiver with a larger support compared to the direct path due to NLOS and antenna effects \cite{UWB-review,gezici-book}. This problem is addressed using techniques such as thresholding \cite{dardari}, using channel information \cite{sahinoglu}, etc. When these algorithms are implemented using digital transceivers, they require sampling rate much higher than the Nyquist rate. Our purpose in this paper, is to develop a method which can operate at lower sampling rate using compressive sampling technique.

IR-UWB based TOA estimation using ML approach is described in \cite{gezici-book}. We will discuss this in greater detail in later section and use this method to compare the proposed schemes in this paper.  When implemented digitally, ML based TOA estimator requires Nyquist-rate sampling. Estimating the TOA in a sparse domain using the received IR-UWB signal  is the main theme of the paper. 

Before we continue the discussion, the main claims of the paper are summarized as follows. 
\begin{itemize}
\item We propose novel compressive sampling algorithms for the IR-UWB based TOA estimation.
\item We propose a compressive sampling dictionary design, so that the received UWB signal can be efficiently expressed as a linear combination of the columns of this dictionary. This results in compact representation of the received UWB signal.
\item We modify the proposed TOA estimation algorithm to exploit the a-priori channel information that may be available.
\item We provide a performance comparison of the proposed methods with ML based TOA estimation methods.
\item We analyze the performance of the methods under practical scenarios for different dictionary parameters and sampling rates.
\end{itemize}
Consequence of these claims indicate that a cost effective  digital UWB TOA estimators can  be developed for the practical UWB applications, thus leading to a progress beyond the state-of-the-art. Besides the achievement over the state-of-the-art, this paper compliments the current trend in the research pertaining to UWB domain including \cite{Wang-TOA, ballal-eui, meng-cs,mao-nonuni}.

\section{Signal Model}
\label{sec:model}
We consider a single user UWB system. The signal model comprises of $N_f$ frames each having an unit energy pulse, $s(t)$, given by
\begin{equation}
  \label{eq:tr}
  \WTR (t)=\sum\limits_{j=0}^{N_f-1}d_js(t-jT_f-c_jT_c),
\end{equation}
where each frame is of duration, $T_f $, and the frame index is represented by $j$. The chip duration is represented by $T_c$ and $c_j \in \{0 \ldots N_c\}$ indicates the time-hopping code. The $d_j\in \{\pm 1\}$ is the polarity code, which can be used along with time-hopping to smooth the signal spectrum. 

The received signal is the distorted version of the transmit pulse with multipaths. The TOA is defined as the time elapsed for the first arrival path to reach the receiver from the transmitter. The received signal can be represented by 
\begin{equation}
  \label{eq:rx}
  \WRX (t)=\sum\limits_{j=0}^{N_f-1}d_jr(t-jT_f-c_jT_c) + n(t),
\end{equation}
where, 
\begin{equation}
  \label{eq:rx1}
r(t)=\sqrt{\frac{E_b}{N_f}}\sum\limits_{\ell=1}^{L}\alpha_\ell s(t-\tau_l).  
\end{equation}
Here, $E_b$, is the captured energy and 
$\sum\limits_{\ell=1}^{L}\alpha_\ell^2=1$. The gain for the $\ell$-th tap is given by $\alpha_\ell$. The $n(t)$ is the AWGN process with zero mean and double-sided power spectral density of $N_0/2$. Without loss of generality, and for simplicity of analysis, we assume $c_j=0$, $N_f=1$, and $d_j=1$. The TOA estimation problem involves estimating the first arrival path, $\tau_1=\ttoa$, in the received signal \eqref{eq:rx}. The frame duration, $T_f$, is chosen sufficiently larger than the delay spread of the channel to avoid any inter pulse interference, that is $T_f\gg T_d$, where, $T_d$, is the delay spread of the channel.

\section{ML based TOA estimation}
 \label{sec:ML}
In order to be self-contained, in this section, a short review of ML based UWB TOA estimation is given \cite{Falsi-toa,gezici-book,Win-commth}. Consider a direct sampling receiver generating the sampled output of \eqref{eq:rx}, defined by $\bb{r}$, such that $r(i)=\WRX(iT_s)$. The $N=T_f/T_s$, is the number of samples corresponding to a frame and $\bb{r}\in\R{N}$. The received samples can be written as
\begin{equation}
  \label{eq:drx}
  \bb{r}=\bb{W}(\bs{\tau})\bs{\alpha} +\bb{n},
\end{equation}
where, $\bs{\alpha}=[\alpha_1, \ldots ,\alpha_L]^{\scriptsize T}$, represents the path-gain,  $\bb{n}\in \R{N}$, are the noise samples with its elements, $n(i)=n(iT_s)$ and $\bb{W}(\bs{\tau})=\left[\bb{w}_{d_1}, \bb{w}_{d_2}, \ldots, \bb{w}_{d_L}\right] \in\R{N\times L}$. The $\bb{w}_{d_i}$, is an $N$ dimensional vector defined as 
\begin{equation}
  \bb{w}_{d_i}=\left[\bb{0}^{\scriptsize T}_{d_i},\bb{w}^{\scriptsize T}, \bb{0}^{\scriptsize T}_{N-P-d_i}\right]^{\scriptsize T}, \nonumber
\end{equation}
where, $\bb{w}$ denotes a vector of discrete samples representing the transmit pulse $s(t)$, with its $i$-th element, $w(i)=s(iT_s)$, $i=0,\ldots,P-1$. The $\bb{0_{d_i}}$ is a zero vector of size, $d_i= \lfloor{\tau_i/T_s}\rfloor$.

The ML estimation for the unknown parameter set $\bs{\nu}=\left[\bs{\alpha}^{\scriptsize T}, \bs{\tau}^{\scriptsize T}\right]^{\mb{\scriptsize T}}$, can be obtained by solving the following optimization problem

\begin{equation}
  \label{eq:mlopt}
  \hat{\bs{\nu}}=\underset{\bs{\nu}}{\argmin} \left\{\frac{1}{N}\norm{\bb{r}-\hat{\bb{r}}}_2^2\right\},
\end{equation}
where the elements of $\hat{\bb{r}}$ are given by
\begin{equation}
  \label{eq:ml1}
\hat{r}(i)=\sum\limits_{\ell=1}^L\hat{\alpha}_\ell s(iT_s-\hat{\tau_\ell}) .
\end{equation}
Solving the optimization \eqref{eq:mlopt} is computationally intensive as it requires a search over entire parameter space $\bb{\bs{\nu}}$. However, if the mutipaths in the channel are separable, then the unknown parameter estimation simplifies to \cite{Falsi-toa,Win-commth}\footnote{Energy of the received pulse is assumed to be one.}

\begin{eqnarray}
  \label{eq:sep}
  \hat{\bs{\tau}}&=& \underset{\bs{\tau}}{\argmax}\left\{ \sum\limits_{i=1}^{L}\left( \bb{w}_{d_i}^{\scriptsize T}\bb{r}  \right)^2\right\}, \\
\hat{\bs{\alpha}}&=&\bb{W}(\hat{\bs{\tau}})^{\scriptsize T}\bb{r}.
\end{eqnarray}

In this case the estimation of the TOA, $\tau_1$, is decoupled from the estimation of the other channel parameters. The optimization of \eqref{eq:sep}, can be accomplished by maximizing each term of the sum independently. We use this method to compare the performance of the proposed algorithms.

\section{UWB Channels}
\label{sec:channel}
One of the most widely used channel models for indoor propagation was proposed by  Saleh and Valenzuela \cite{Saleh-Channel}. This model is adopted in IEEE 802.15.4a   CM1 standard for providing stochastic channel model for LOS residential conditions \cite{uwb-channel-model,Molisch}. As per this model, the  discrete-time impulse-response of the UWB channel has clusters arriving in the Poisson distributed way and the multipath components (MPCs) within the cluster follows a Laplacian model. A  model with $C$ clusters having $R$ rays (MPCs) can be expressed as
\begin{equation}
  \label{eq:saleh}
  h(t)=\sum\limits_{i=1}^{C}\sum_{j=1}^{R}a_{i,j}\delta(t-T_i-\gamma_{\tiny i,j}),
\end{equation}
where, $T_i$ represents the arrival time of the $i$-th cluster and  $\gamma_{i,j}$ represents the $j$-th ray in the $i$-th cluster. In \eqref{eq:rx1}, we have combined the cluster and ray arrivals, such that
\begin{eqnarray*}
  \label{eq:s1}
  \left[\alpha_1,\ldots,\alpha_L\right]&=&\left[a_{\tiny 1,1},\ldots, a_{\tiny C,R}\right], \\
\left[\tau_1,\ldots,\tau_L\right]&=&\left[(T_1+\gamma_{\tiny 1,1}), \ldots, (T_C+\gamma_{\tiny C,R}) \right],
\end{eqnarray*}
where, $L=CR$. Note that by definition $\gamma_{i,1}=0$. Therefore, $T_1=\tau_1=\ttoa$, denotes the arrival time of first ray  of the first cluster and is the TOA for the LOS UWB channel. For details about the cluster and ray arrival rates, refer to the IEEE 802.15.4a CM1 model described in \cite{802.15.4a}.

\begin{figure}[t]
\centering
\includegraphics[width=0.7\textwidth]{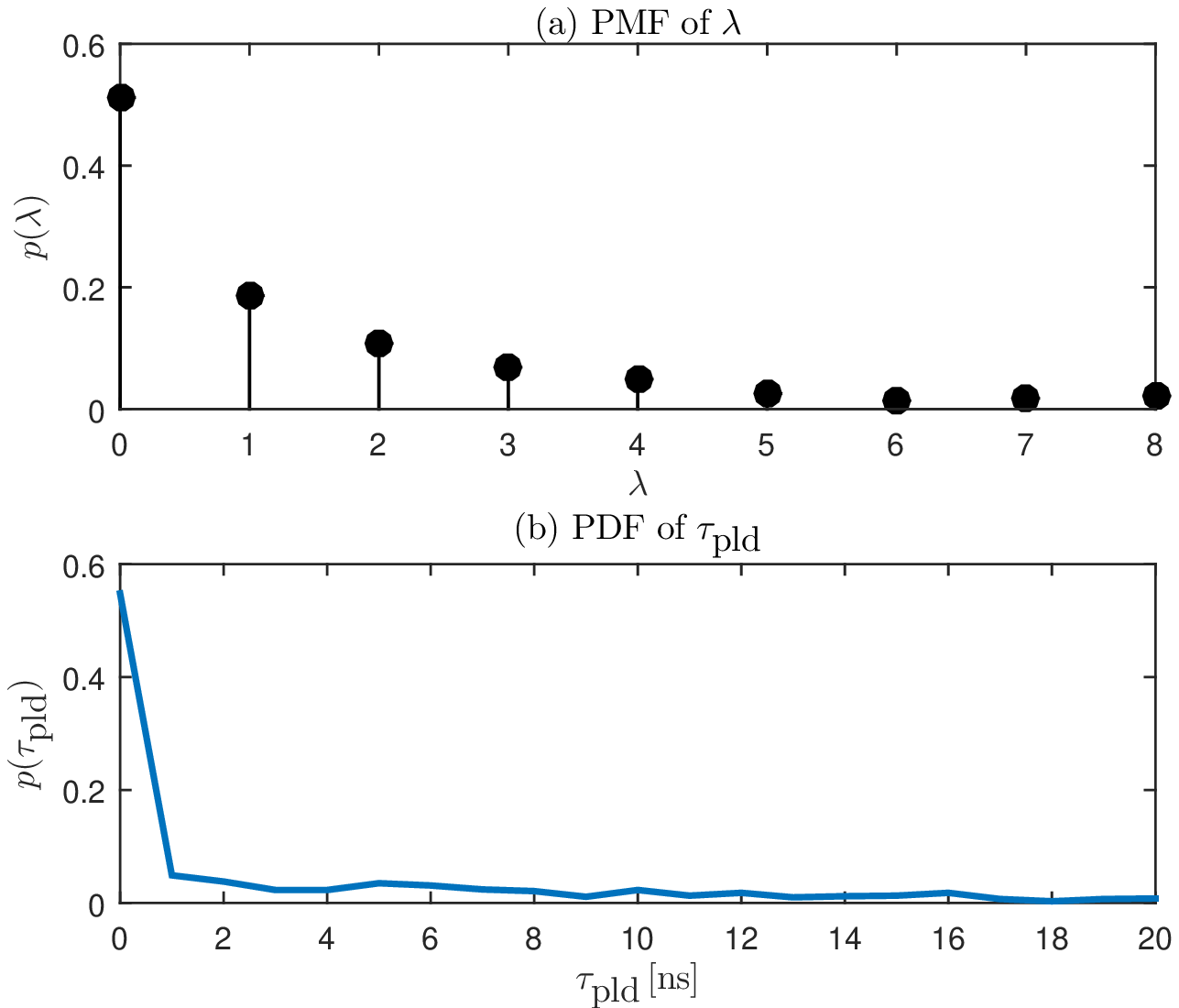}
\caption{The PMF and PDF of $\lambda$ and $\pld$ respectively for IEEE 802.15.4a CM1 model. The sampling frequency of $8~\mb{GHz}$ and $100$ distinct channel realization are employed.}
\label{fig:APR}
\end{figure}

Below, we will derive  new statistical parameters for the IEEE 802.15.4.a CM1 model, which can be utilized by the TOA estimation algorithm to improve the performance. For a multipath  residential LOS channel proposed in IEEE 802.15.4a CM1 model, if we consider all the significant paths that constitutes $80\%$ of the total energy, then the probability mass function (PMF) for the number of significant paths, $\lambda$,  that arrives before the strongest path is shown in the Fig.~\ref{fig:APR}(a). Approximately $50\%$ of the time the first arriving path is weaker than the strongest path. If we define $\pld$ as the peak to first path delay, that is,   $\pld=\tau_{\mb{\scriptsize peak}}-\tau_1$, where, $\tau_{\mb{\scriptsize peak}}$, is the location of the peak, then the probability density function of the $\pld$ is shown in Fig.\ref{fig:APR}(b).

Channel  a-priori information shown in Fig.~\ref{fig:APR}(a) and Fig.~\ref{fig:APR}(b) combined with the geographic constraint on the range can be used to improve the performance of the proposed compressive sampling algorithms. We will discuss this in the later section. In the next section, we will briefly introduce compressive sampling theory and discuss the representation of the received UWB signal, $\bb{r}$,  in a sparse domain.

\section{Sparse representation of UWB signal}
\label{sec:cs}

\begin{figure*}[t]
  \centering
  \includegraphics[scale=0.6]{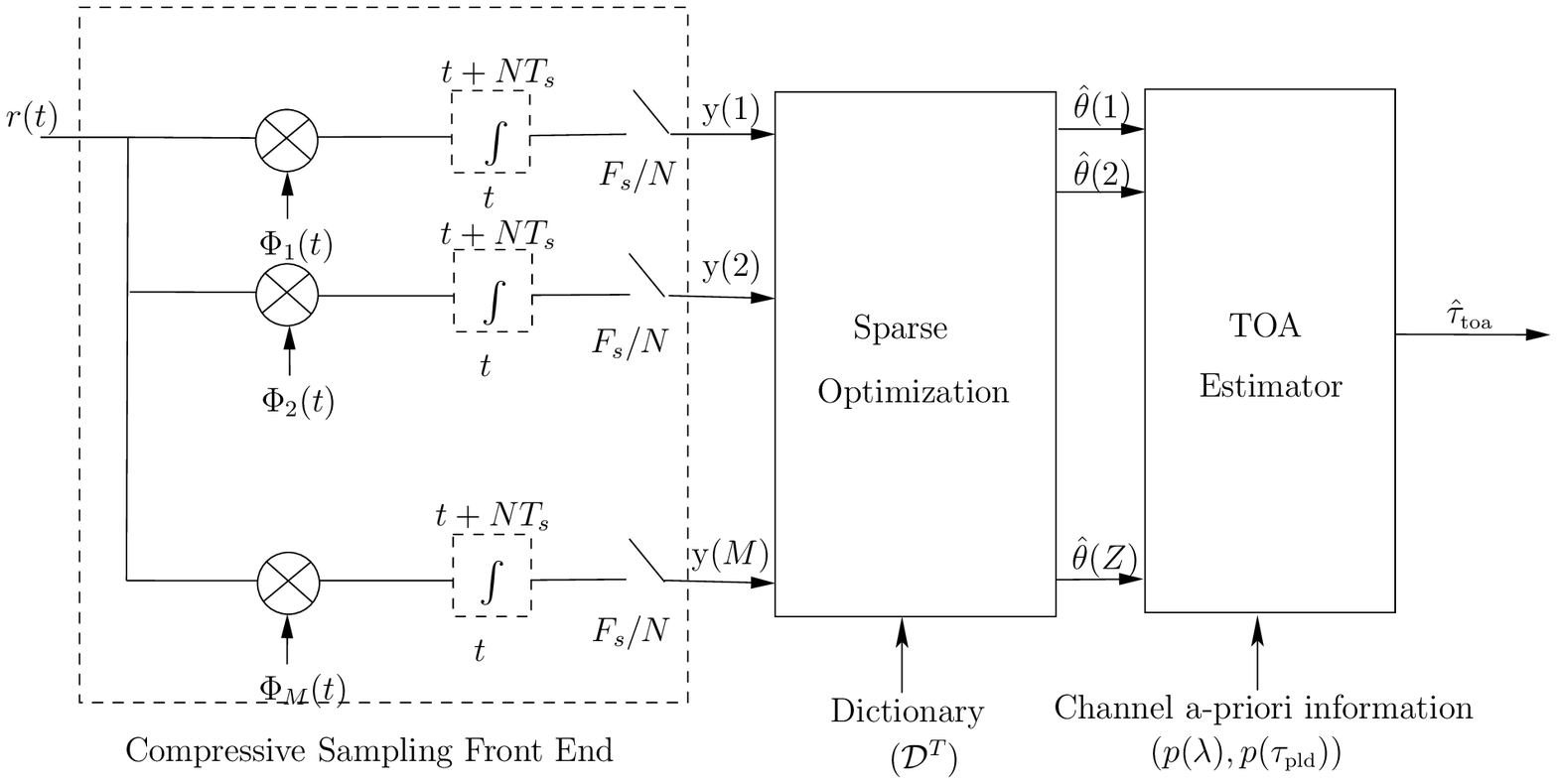}
  \caption{Block diagram of the compressive sampling (CS) system for TOA estimation. The $\Phi_i(t),i=1,\ldots,M$, denotes the continuous time i.i.d normal processes with zero mean and unit variance.}
  \label{fig:CSHardware}
\end{figure*}

Consider an $N$-point Nyquist sampled discrete-time representation of the received UWB signal, $\mbf{r}\in \R{N}$ obtained by sampling the received signal, $r(t)$ using an ADC at rate $F_s$. We propose to modify the signal acquisition hardware as shown in Fig.~\ref{fig:CSHardware}, where the signal is acquired at a lower sampling rate, ($F_s/\mathcal{U}$), where, $\mathcal{U}=N/M$, is the under-sampling ratio. The acquired signal, $\mbf{y}$, can be viewed as a projection of $\mbf{r}$ on a measurement matrix, $\mbs{\Phi}$. The elements of $\mbs{\Phi}$ are i.i.ds drawn from a normal distribution. That is,

\begin{equation}
  \label{eq:s_pr}
  \mbf{y}=\mbs{\Phi}\mbf{r},
\end{equation}
and $\mbs{\Phi}$ is a $M \times N$ matrix,  with $M=N/\mathcal{U}$. As shown in Fig.~\ref{fig:CSHardware}, $\mbf{y}\in \R{M}$, is fed to the sparse optimization routine, which will represent  $\mbf{y}$ in a compact form using the columns of a carefully chosen dictionary leading to a sparser signal representation. 

We consider a dictionary,
$\BD^T$, whose dimension is  $N \times Z$,  whose columns, $\mbf{d_i}$, are the basis vectors in space $\R{N}$, in which, $\mbf{r}$ is sparse, that is 
\begin{equation}
  \label{eq:s_dict}
  \mbf{r}=\sum\limits_{i=0}^{S}\theta_{\ell_i}\mbf{d_i},
\end{equation}
where $S \ll Z$,  is the sparsity of the received UWB signal in the dictionary domain. The $\ell_i$ is the support of vector $\mbs{\theta}\in\R{Z}$ and $\theta_{\ell_i}$  denotes the non-zero value at the support. The sparse optimization block of Fig.~\ref{fig:CSHardware}, typically solves an optimization problem to represent $\mbf{y}$, using a sparse vector $\hat{\bs{\theta}}\in\R{Z}$. For example, a basis pursuit type of algorithm solves,
\begin{align}
  \label{eq:opt}
\mbs{\hat{\theta}}=&\argmin \norm{\mbs{\theta}}_1, \\
&\text{subject to } \mbf{y}=\HH \bt,  \nonumber 
\end{align}
where, $\mbs{\mathcal{H}}=\mbs{\Phi}\BD$, is called holographic dictionary. From the sparse representation, an estimate of the TOA, $\httoa$ needs to be estimated. 

In the reminder of the paper,   for the proposed structure in Fig.~\ref{fig:CSHardware}, we will discuss how to design a compressive sampling dictionary. Then we will propose a TOA estimation algorithm, which will utilize the sparse representation of the received UWB signal in the chosen dictionary domain to estimate the TOA. We will modify the algorithm to utilize the a-priori information regarding the channel and the geographical constraints to improve the performance of the proposed TOA estimation algorithm. 

\section{Method for sub-Nyquist TOA Estimation}
\label{sec:algo}
In this section, first, in Section \ref{ss:dict}, we design the compressive sampling dictionary for the TOA algorithms. Subsequently, two main contributions of the paper which uses the above designed dictionary for TOA estimation are discussed in some detail. In Section \ref{ss:no-apr}, we discuss the TOA estimation using a modified greedy search algorithm, then the TOA estimation based on a-priori channel information is studied in Section \ref{ss:apr}.

\subsection{Compressive sampling dictionary}
\label{ss:dict}
As shown in \eqref{eq:rx}, the received UWB signal is a scaled, and delayed version of the transmit pulse. The transmit pulse, $s(t)$, is typically chosen as first or second order Gaussian derivative pulse {\cite{VJ-3}. To construct the dictionary, we choose each column of the dictionary, $\BC^T(t,\Delta)$, as shifted versions of the transmit pulse, that is
\begin{equation}
  \label{eq:d2}
  \BC^T(t,\Delta)=\left[p_0(t),\cdots,p_{Z-1}(t)\right],
\end{equation}
where
\begin{equation}
  \label{eq:d1} 
p_\ell(t)=s(t- \Delta \ell), \mb{ } \ell=0,2,\cdots,Z-1.
\end{equation}
The $Z$, defines the number of atoms (columns)  in the dictionary.  The offset, $\Delta$, needs to be controlled to strike a compromise between the number of atoms needed to faithfully represent the received UWB pulse and the size of the dictionary. The $\Delta$ and $Z$ are chosen such that $T_f=Z\Delta$. Equation \eqref{eq:d1}, is expressed in terms of continuous $t$ and $\Delta$. In practice, both these parameters are discretized such that, $t$, is sampled at a particular sampling period, $T_s$, and $\Delta$, is a multiple of $T_s$. We define $\BD$ as uniform sampling of the dictionary $\BC$, and is of dimension $Z \times N $ as each atom ($p_\ell(t)$) in the dictionary is now a vector of length $N$. That is 
\begin{equation}
  \label{eq:c-dict}
  \BD^{\scriptsize T}=\left[\bb{w}_0,\ldots,\bb{w}_Z \right] \in \R{N\times Z}, 
\end{equation}
where,
\begin{equation}
  \label{eq:w}
  \bb{w}_\ell=\left[\bb{0}_{\ell N_0},\bb{w},\bb{0}_{N-P-\ell N_0}\right]^{\mb{\scriptsize T}} \in \R{N},
\end{equation}
with $\ell=[0,\ldots,Z-1]$ and $\Delta=N_0T_s$, where, $N_0$, is an integer constant, which can be used to control the size of the dictionary.

\subsection{TOA estimation algorithm}
\label{ss:no-apr}

Consider discrete samples of the received UWB signal  sampled at $F_s$ represented as a time-domain vector,  $\mbf{r}$. Let $\mbf{y}=\Phi\mbf{r}$, denote the random projection of $\mbf{r}$ on the measurement matrix, $\Phi$ (refer to Fig.\ref{fig:CSHardware}), where, $\Phi$, is a $M \times N$ matrix with its elements $\phi_{i,j}$ drawn from $\mathcal{N}(0,1)$.  As shown in Fig.~\ref{fig:APR}(a), for a LOS UWB channel the first arrival path can be weaker than the strongest path. To locate the true TOA,  one has to search backward from the peak location to locate any possible significant energy paths, which may not be strongest. To accomplish this in sub-Nyquist domain, we propose a TOA algorithm as shown in Algorithm \ref{alg:toa}. 

Input to the Algorithm  is the vector $\mbf{y}$, Holographic dictionary, $\HH$, and the parameter, $K$, which defines the number of paths to be searched. In line-$7$$, \HH_{I_k}$,  denotes the matrix composed of the columns defined in the set $I_k$.  The $\HH_{I_k}^{\dagger}$ indicates the pseudo inverse of $\HH_{I_k}$. Also, all columns of $\HH$, are normalized, that is $\norm{\mbf{h_i}}_{2}^{2}=1, \forall i \in[1,\cdots,Z]$, where $\mbf{h_i}$ denotes the $i$-th column of the matrix $\HH$. It can be noticed that in each iteration, the holographic dictionary, $\HH$, is searched for the strongest delayed version of the transmitted signal, that is contained in the residual signal, $\be{k}$. After $K$ iteration, the lowest indexed column of the holographic dictionary, $\ell$, is identified. Since each column in the dictionary is offset by $\Delta$, the TOA is estimated as $\httoa=\ell\Delta$.

The performance of the algorithm depends on the offset, $\Delta$, if $\Delta$ is small, the accuracy of the TOA estimation will be better, as the atoms of the dictionary can resolve the TOA better. However,  this will increase  the dictionary size, $Z$ as $Z=T_f/\Delta$, thereby increasing the memory requirements for the system. 

If the number of paths searched, $K$, in the algorithm is too high then, there is a potential problem of picking the wrong atom in the dictionary due to the noise and if $K$ is too small then there is a possibility of missing the atom which correspond to the weaker first path. There exists an optimal $K$, at which the performance of the estimator is maximum. 

The TOA estimation performance can be  improved by increasing the number of random projections of UWB signal, since it aids better reconstruction \cite{Donoho,Candes}. However, this leads to higher sampling rate and increases the demands on ADC resources, there by increasing the cost of the transceiver as discussed in Section~\ref{sec:intro}. In Section \ref{sec:res}, the performance of Algorithm \ref{alg:toa} is studied as well as the rules-of-thumb is provided for its configuration.

\subsection{TOA estimation with a-priori information}
\label{ss:apr}
If we know certain statistical properties of the channel beforehand, then we can exploit this information to improve the performance of the TOA estimator. For example, for IEEE 802.15.4a CM1 channel model, the probability distributions of number of significant paths before peak-path, $\lambda$, and peak-to-first path delay, $\pld$, are as shown in Fig.\ref{fig:APR}(a) and Fig.\ref{fig:APR}(b). From this we can notice that more than $50\%$ of the time the peak path is the first path, and also probability that the first path is more than $20\,\mb{ns}$ away from the peak location is negligible. Another important a-priori information could be the from the geographic constraints on the range of the target, resulting in TOA values being  $\ttoa < \tmax$.

The above a-priori information can be handled by modifying  Algorithm \ref{alg:toa}, such that, only the paths within a window interval before the peak are considered for the TOA estimation. The modified algorithm is as shown in Algorithm. \ref{alg:toa_apr}. In the later section, we will show that the modified algorithm with a-priori information outperforms the Algorithm~\ref{alg:toa}, which is agnostic to this information.

In the next section, we will assess the performance of the proposed algorithm  in simulations.

\begin{algorithm}[t]
\DontPrintSemicolon 
\caption{TOA estimation algorithm}
\label{alg:toa}
\small
\KwIn{$\mbf{y}, \HH , K, \Delta, \be{0} = \mbf{y}, k=0, I_k= \{\emptyset \} $ }
\KwOut{TOA estimate, $\httoa$.}
\Repeat {$k \le K$} {
$t=\argmax |\langle \bh{i},\be{k} \rangle|, i=1,\cdots,Z$ \; 
\lIf{k==0} { $\ell=t$;}
\lElseIf{t<l} { $\ell=t$; }
$k=k+1$\;
$I_k=I_k\cup t$ \;
$\be{k}=\mbf{y}-\HH_{I_k}\HH_{I_k}^{\dagger}\mbf{y}$
}
\Return{$\httoa=\ell\Delta$}\;
\end{algorithm}

\begin{algorithm}[t]
\DontPrintSemicolon 
\caption{TOA estimation algorithm (with a-priori information)}
\label{alg:toa_apr}
\small
\KwIn{$\mbf{y}, \HH , K, \Delta, \be{0} = \mbf{y}, k=0, I_k= \{\emptyset \}, \tmax, \pldmax $ }
\KwOut{TOA estimate, $\httoa$.}
\tcp{Search only in $\lfloor{\tmax/\Delta}\rfloor$ columns of $\HH$}
$\HH$=$\HH(:,1:\lfloor{\tmax/\Delta}\rfloor)$\;
\tcp{Bound the first path in relation to peak}
$\Omega=\lfloor{\pldmax/\Delta}\rfloor$\;
\Repeat {$k \le K$} {
$t=\argmax |\langle \bh{i},\be{k} \rangle|, i=1,\cdots,\lfloor{\tmax/\Delta}\rfloor$ \; 
\lIf{k==0} { $\ell=t$;}
\lElseIf{t<l and t>l+$\Omega$} { $\ell=t$; }
$k=k+1$\;
$I_k=I_k\cup t$ \;
$\be{k}=\mbf{y}-\HH_{I_k}\HH_{I_k}^{\dagger}\mbf{y}$
}
\Return{$\httoa=\ell\Delta$}\;
\end{algorithm}

\begin{figure*}[t]
  \centering
  \includegraphics[scale=0.6]{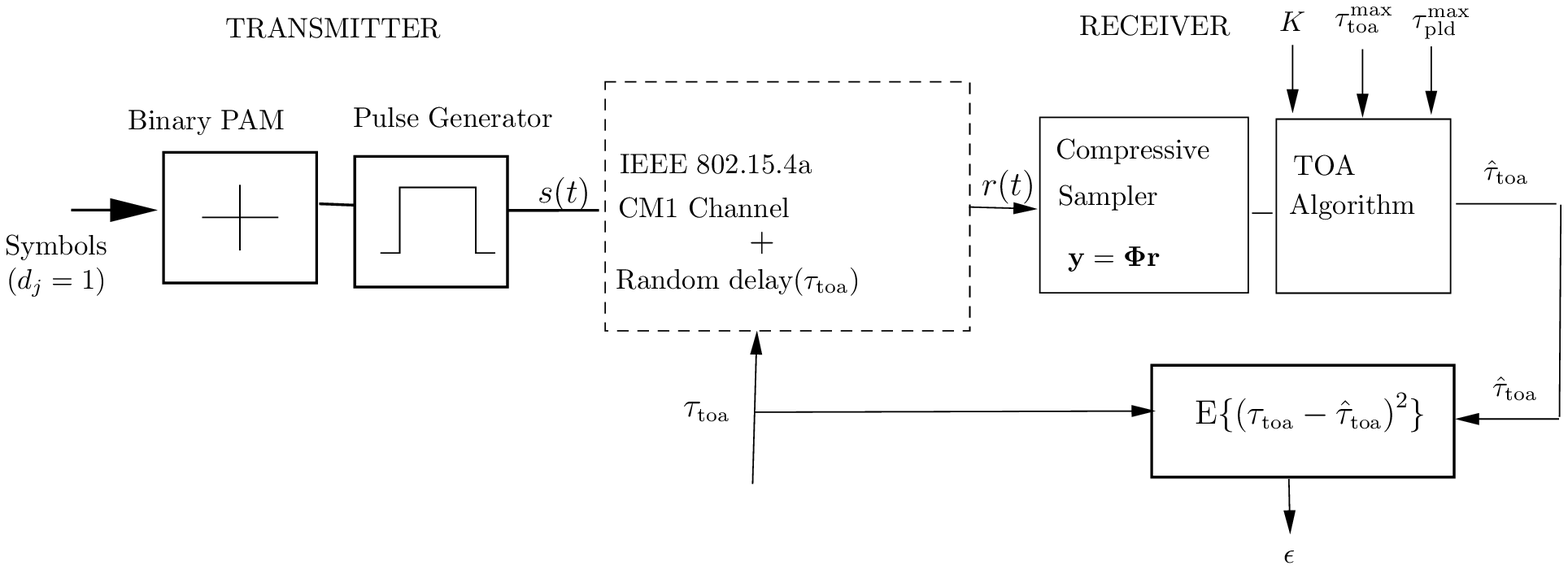}
  \caption{The block diagram of the simulation bed  for the performance evaluation of TOA estimation algorithms.}
  \label{fig:SimulationBed}
\end{figure*}
\begin{figure}[t]
  \centering
   \includegraphics[width=0.7\textwidth]{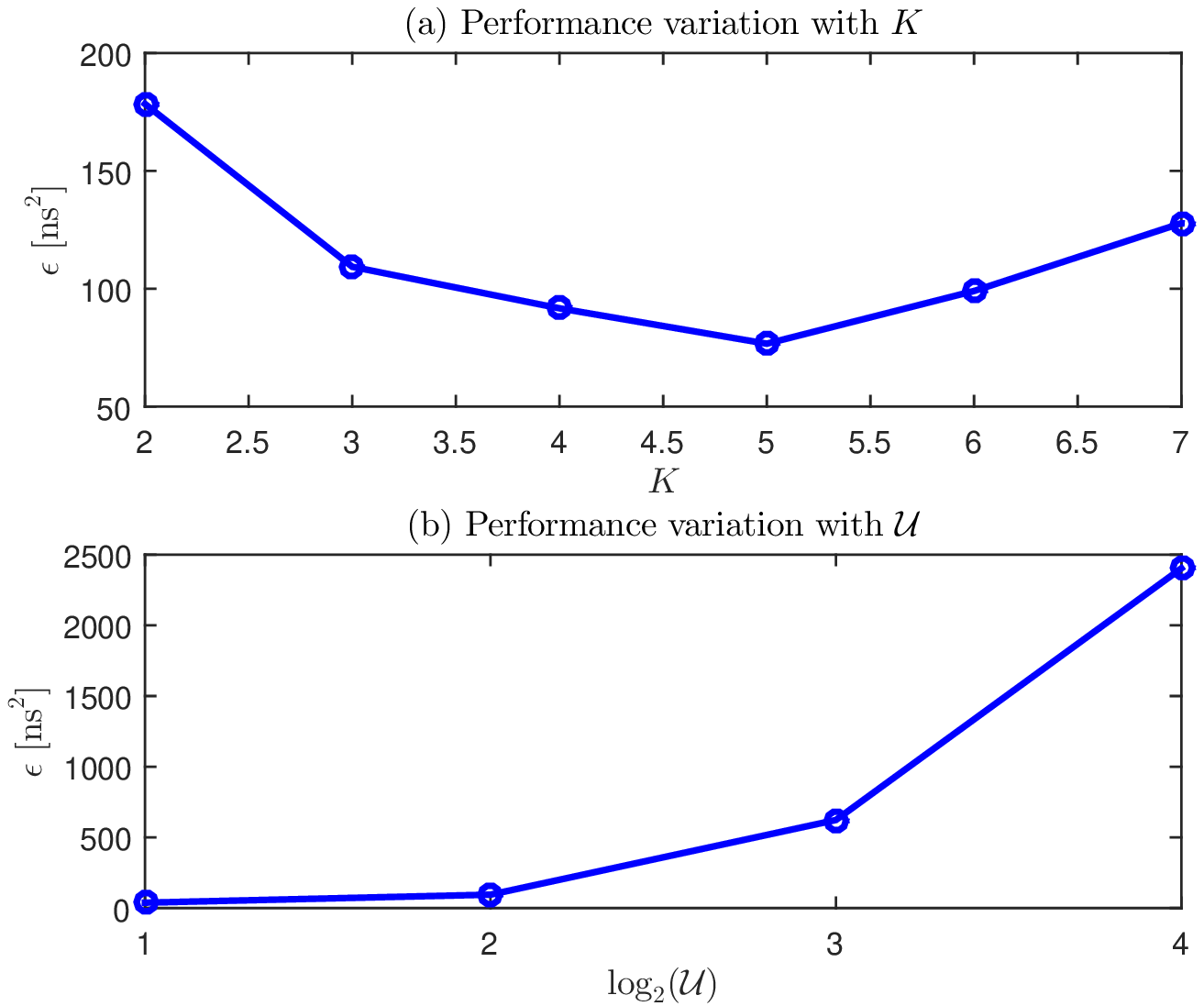}
  \caption{The performance of the proposed method with number of paths searched, $K$, and under-sampling ratio, $\mathcal{U}$, for a fixed $\mb{SNR}=24~\mb{dB}$. $\mathcal{U}=N/M=4$, and $K=5$, is employed in (a) and (b) respectively.}
  \label{fig:KU_sweep}
\end{figure}

\section{Simulation Study}
\label{sec:res}
In this section, the performance of the proposed methods are investigated by numerical simulations that mimic a realistic UWB link. Residential LOS UWB channels can vary depending on the environmental  aspects such as plan of the building, type of walls, obstacles, etc. The IEEE 802.15.4a UWB stochastic channel models, which are developed based on actual measurements from the measurement setup described in \cite{uwb-channel-model,Molisch} represents the practical UWB channels.  Based on this channel model, a simulation bed for a single user UWB system is developed for the evaluation of the proposed TOA algorithms. The block diagram of this is as shown in the Fig.~\ref{fig:SimulationBed}. We choose  $F_s=8~\mb{GHz}$, $T_f=200~\mb{ ns}$, and $N_f=1$. We used a second order Gaussian pulse of width of $1$~\mb{ns} as defined in \cite{gezici-book} as the transmit pulse. Thousand distinct TOA values, drawn from a uniform distribution of $[0~\mb{ns} - 50~\mb{ns}]$, is employed in the simulation. The TOA modulated pulses are received using $1000$, distinct  realizations of IEEE 802.15.4a CM1 channel. We use the mean square error (MSE), ${\epsilon=\mbf{E}[(\ttoa-\httoa)^2]}$, as the metric to assess performance.  Here, $\ttoa$, is the true TOA and $\httoa$, is the estimated TOA. To give a full picture, we compare the performance of the proposed algorithm with the ML estimation method described in Section~\ref{sec:ML}. The optimization described in \eqref{eq:sep} is evaluated, with $L=10$ and $F_s=8~\mb{GHz}$.

\begin{figure}[t]
  \centering
  \includegraphics[width=0.7\textwidth]{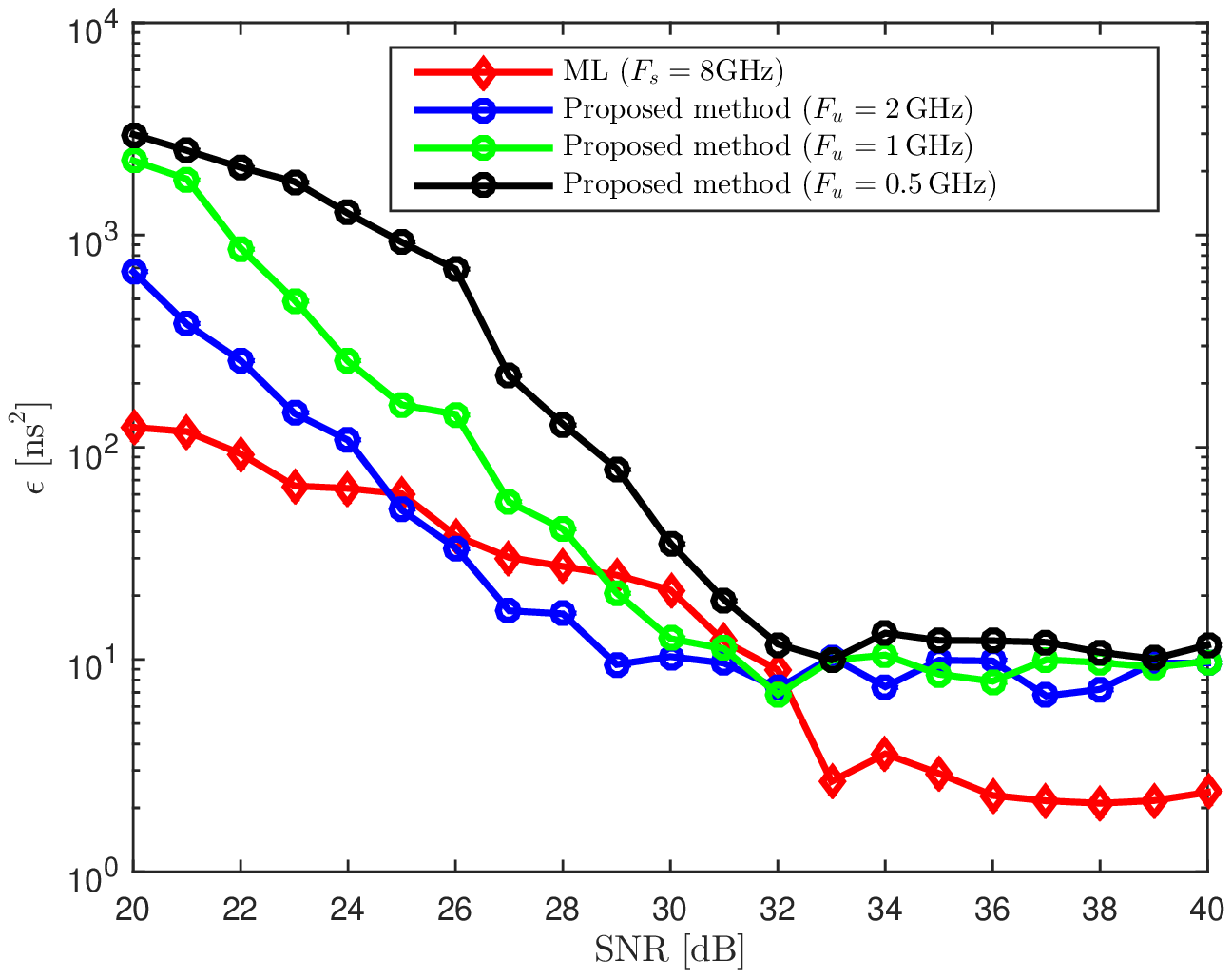}
  \caption{The Performance of the proposed and ML based TOA estimation algorithms.  The second order Gaussian pulse of width of $1$~\mb{ns} is employed in the simulation and $\Delta=T_s$ is considered for the dictionary construction. }
  \label{fig:F_sweep}
\end{figure}

\begin{figure}[t]
  \centering
  \includegraphics[width=0.7\textwidth]{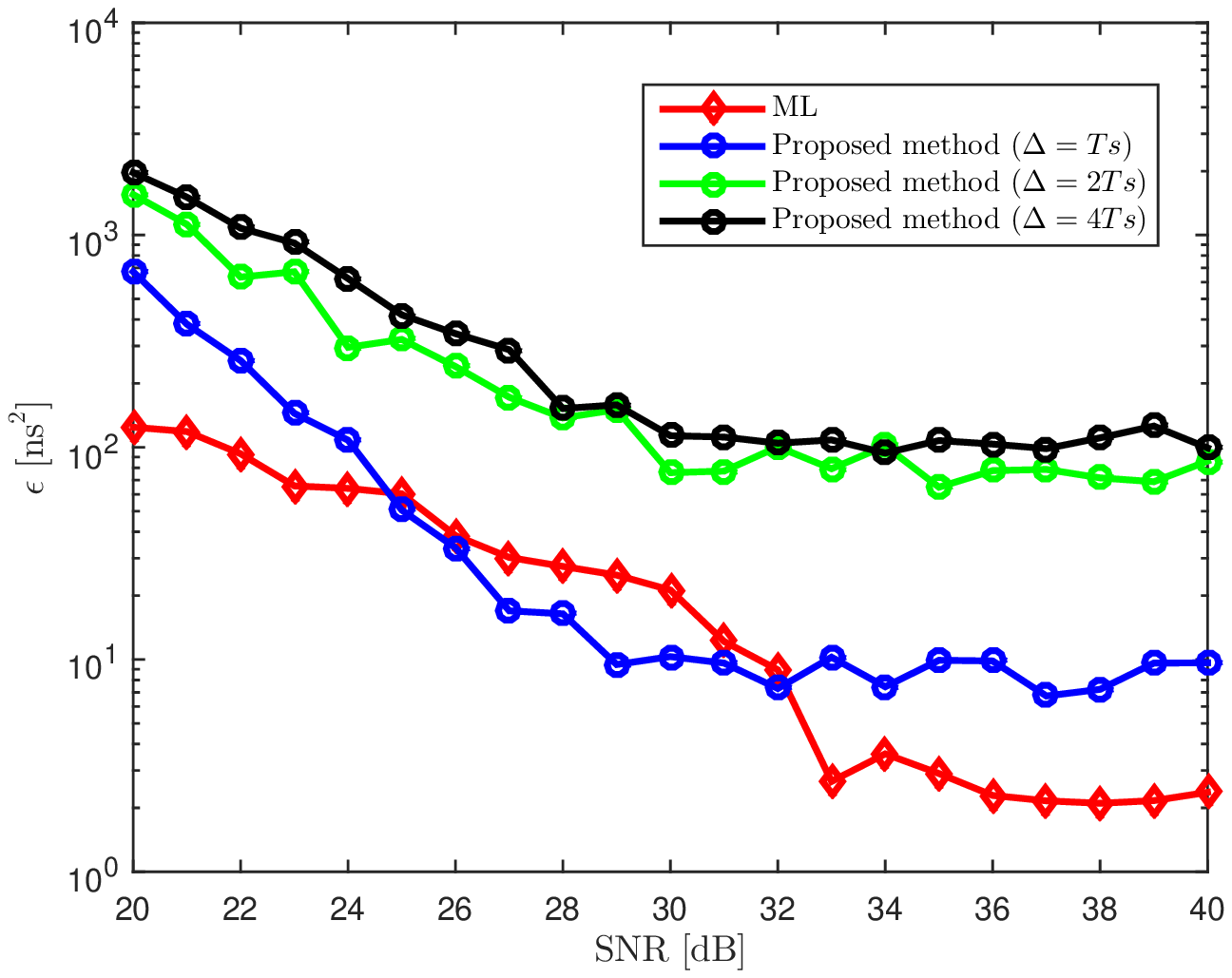}
  \caption{Performance of the proposed and ML based  TOA estimation algorithms.  The ML algorithm with sampling frequency, $F_s=8~\mb{GHz}$, and proposed method with under-sampling ratio, $\mathcal{U}=4$, \emph{i.e.}, ${F_u=F_s/\mathcal{U}=2~\mb{GHz}}$, are compared with different $\Delta$s. }
  \label{fig:SNR_Offset_sweep}
\end{figure}

\begin{figure}[t]
  \centering
  \includegraphics[width=0.7\textwidth]{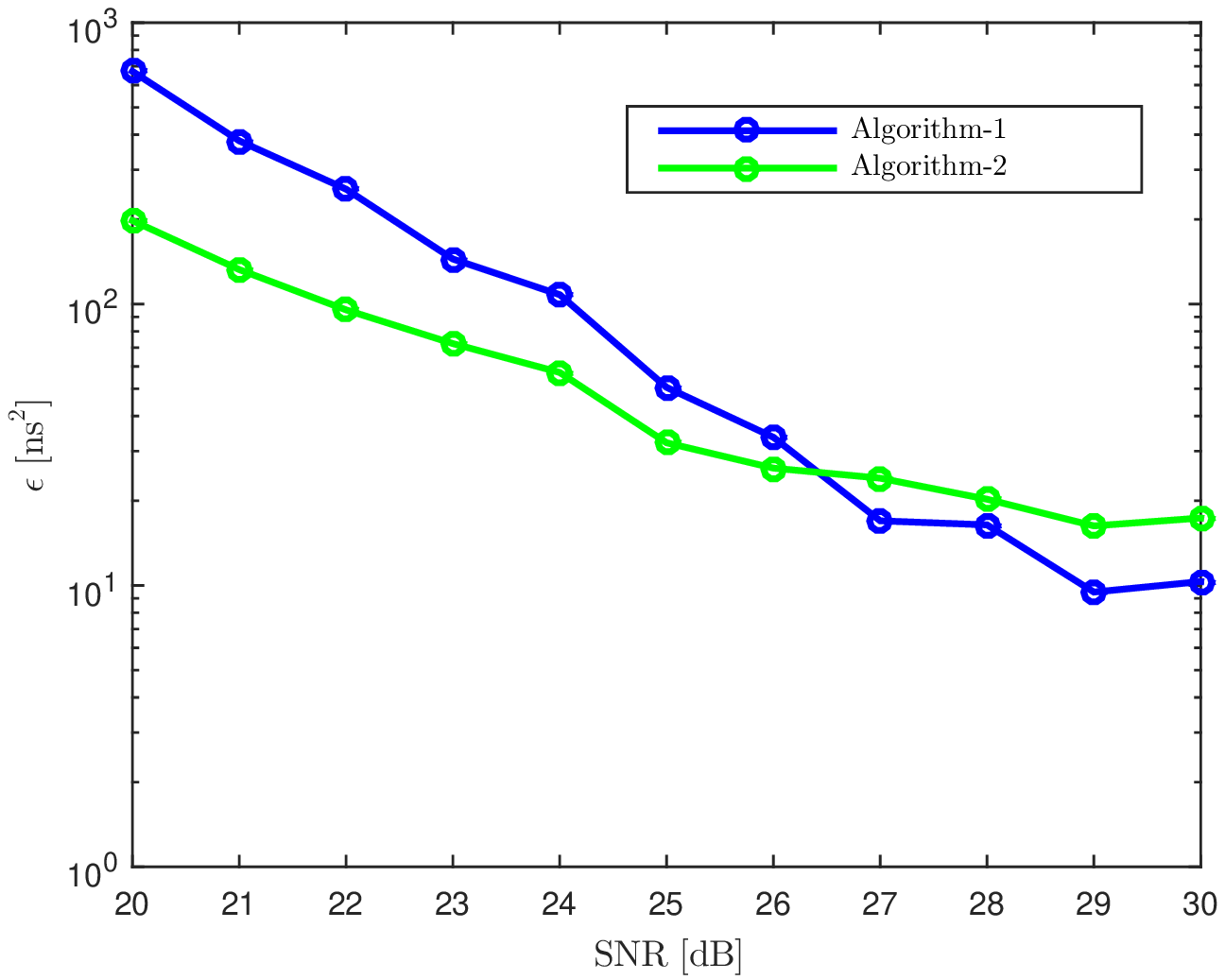}
  \caption{Performance comparison of Algorithm \ref{alg:toa_apr} and Algorithm \ref{alg:toa} in the presence of a-priori information such as $\pldmax$ and $\tmax$. In the Algorithm \ref{alg:toa_apr}, we exploited a-priori information that $\tmax=50\,\mb{ns}$ and from the probability distribution of  $\pldmax$ for IEEE 802.15.4a channel model (refer to Fig.~\ref{fig:APR}(b)), we set $\pldmax=20\,\mb{ns}$.   }
  \label{fig:SNR_Apr_Offset_sweep}
\end{figure}

\subsection{Choice of $K$ and $\Delta$}
\label{ss:kselection}

As discussed earlier, the choice of $K$ and $\Delta$ plays a significant role in the performance of the algorithm. If the number of paths searched, $K$, in the algorithm is high then there is a potential problem of picking the wrong atom (column) in the dictionary due to the noise, however, if the $K$, is too small then we may miss the true TOA, due to the possibility of earlier paths being weaker than the first path.  Typically $K$ can be selected by solving the optimization problem 
\begin{equation}
  \label{eq:optK}
  \underset{K}{\argmin}\mb{ E}\{(\ttoa-\httoa)^2\},
\end{equation}
for a fixed, $\mathcal{U}$, SNR. For a LOS UWB channel models like IEEE 802.15.4a model, it is difficult to arrive at the closed form equation. We measure MSE, $\epsilon$, for various value of $K$ as shown in Fig.~\ref{fig:KU_sweep}a, to pick optimal $K$.

The performance of the proposed algorithm depends on the resolution, $\Delta$, of the dictionary. This is illustrated in the Fig.~\ref{fig:KU_sweep}b. The smaller resolution yields better performance, however this requires larger dictionary size.

\subsection{Performance with no a-priori information}
\label{ss:sim_noapr}
Below, we present one of the main result of this paper. The performance of the Algorithm~\ref{alg:toa} is evaluated for various SNRs, under different  practical settings such as dictionary size, sampling frequency, etc. We use the simulation setup as discussed in the beginning of this section. Note that  Algorithm~\ref{alg:toa} does not assume any prior knowledge regarding the TOA range or the UWB channel.

In Fig.~\ref{fig:F_sweep}, the  performance of the algorithm is evaluated for various under-sampling ratio, $\mathcal{U}$. As described above second order Gaussian pulse of width of $1$~\mb{ns} and $\Delta=T_s$ are considered for the dictionary construction. As expected, the performance degrades with the decrease of the sampling rate $F_u=Fs/\mathcal{U}$.

In the Fig.~\ref{fig:SNR_Offset_sweep}, the  performance of the algorithm is compared with the ML estimation method described in Section \ref{sec:ML} for various $\Delta$s at under-sampling ratio, $\mathcal{U}=4$. As described in Section  \ref{ss:dict}, the larger $\Delta$, will ease the memory requirement of the TOA system due to the reduced  dictionary size, however, this will result in inferior performance as shown in Fig.~\ref{fig:SNR_Offset_sweep}. 

The plots indicate that the proposed algorithm can yield similar performance as ML estimation, with only fraction of the sampling rate at high SNRs.  Based on the environment, engineering  trade offs between, $\Delta$, $\mathcal{U}$ and $K$  need to be done while employing the proposed compressive sampling TOA algorithm for TOA estimation.

\subsection{Performance with a-priori information}
\label{ss:sim_apr}
We use the same setup for the simulation as described in the beginning of this section. Notice that in our simulation setup the TOA ranges are uniformly distributed between  $[0~\mb{ns} - 50~\mb{ns}]$. We also know that the channel model employed is IEEE 802.15.4a CM1 model. From Section \ref{sec:channel}, we notice that for this channel the probability that $\pld>20\,\mb{ns}$ is negligible. 

We evaluates the TOA estimation performance using Algorithm~\ref{alg:toa_apr} with $\tmax$ set to $50\,\mb{ns}$ and $\pldmax$ set to $20\,\mb{ns}$. The variation of MSE, $\epsilon$, with SNR is given in Fig.~\ref{fig:SNR_Apr_Offset_sweep}. Notice that by exploiting the channel and geographical constraints performance of the estimation at low SNR can be significantly improved.

To summarize,  we have shown that the proposed TOA estimation algorithm, together with the proposed compressive sampling dictionary,  can achieve performance comparable to ML algorithm with only a fraction of the sampling frequency at high SNRs. TOA algorithms require a reasonable range to error  ratio for it to be viable for many applications, having a slight loss of performance (\textless 5\%), compared to ML estimator with a significant savings in the ADC resources make the proposed methods pertinent for many applications. At low SNRs, the performance of the proposed method can be improved by  exploiting the a-priori information.

\section{Conclusions}
\label{sec:con}

A compressive sampling based TOA algorithm is proposed in this paper. The proposed algorithm along with the proposed dictionary can yield the same performance as the ML based TOA estimation with only $1/4$-th the sampling rate at $25~\mb{dB}$ SNR as shown in Fig.~\ref{fig:SNR_Offset_sweep} for IEEE 802.15.4a CM1 channel model. We also analyzed, how the performance of the algorithm varies with the choice $K$ and $\Delta$. This is shown in Fig.~\ref{fig:KU_sweep}. From Fig.~\ref{fig:KU_sweep}a, we notice that there exists an optimal $K$, which maximizes the performance for the choice of algorithm parameters. The trade off between the size of the dictionary and the performance is shown in  Fig.~\ref{fig:KU_sweep}b. Impact of the sampling rate on the performance is shown in the Fig.~\ref{fig:F_sweep}. As expected, the performance of the proposed method degrades with the decrease of the sampling rate, $F_u=F_s/\mathcal{U}$.

In certain TOA estimation scenarios, where the a-priori information is available then the performance of the proposed algorithm can be improved. In many scenarios this can originate from the geographical constraints on the range of the target which can upper bound the TOA so that $\ttoa < \tmax$ or the channel model from which we can limit the search interval by exploiting peak-to-first path lag, $\pldmax$. We modified the Algorithm~\ref{alg:toa} to Algorithm~\ref{alg:toa_apr} to exploit these a-priori information and demonstrated the benefits through Fig.~\ref{fig:SNR_Apr_Offset_sweep}.

The results are encouraging for the UWB TOA estimators working in the high SNR scenarios. The consumption of ADC resources can be significantly reduced thereby reducing the cost of the transceiver. In a typical UWB residential LOS channels, high SNRs in excess of $25~\mb{dB}$ can be achieved by averaging over many independent  received frames. The proposed TOA estimation method  is sub-optimal as it does not take in to consideration the frequency selective characteristics  of the UWB antenna. We intend to further develop the work to address these practical issues.

\begin{backmatter}

\section*{Competing interests}
  The authors declare that they have no competing interests.

\section*{Acknowledgements}
Parts of the work have been funded by The Swedish Agency for Innovation Systems (VINNOVA).


\bibliographystyle{bmc-mathphys} 
\bibliography{my}      




\end{backmatter}

\end{document}